\documentclass[aip,graphicx]{revtex4-1}

\draft % marks overfull lines with a black rule on the right

\begin{document}

\title{Covariant diagonalization of the perfect fluid stress-energy tensor} %Title of paper

\author{Alcides Garat}
%\thanks{}
\affiliation{1. Instituto de F\'{\i}sica, Facultad de Ciencias, Igu\'a 4225, esq. Mataojo, Montevideo, Uruguay.}

\date{April 14th, 2012}

\begin{abstract}
We introduce new tetrads that manifestly and covariantly diagonalize the stress-energy tensor for a perfect fluid with vorticity at every spacetime point. This new tetrad can be applied and introduce simplification in the analysis of astrophysical relativistic problems where vorticity is present through the Carter-Lichnerowicz equation. This new tetrad also enables the construction in a simple fashion of Euler and Coordinate observers relevant to the Cauchy evolution of many hydrodynamical relativistic problems in the case with no vorticity and the presence of a symmetry where the tetrads are completely analogous to the case with vorticity. We also discuss the origin of inertia in this special case from the standpoint of our new local tetrads.
\end{abstract}

\pacs{}% insert suggested PACS numbers in braces on next line

\maketitle %\maketitle must follow title, authors, abstract and \pacs

\section{Introduction}
\label{intro}

Relativistic fluid dynamics is a subject of relevance in three main astrophysical problems as it emerges, for instance, from the analysis in references \cite{EG}$^{,}$\cite{CW}$^{,}$\cite{AC}$^{,}$\cite{F}$^{,}$\cite{NS}$^{,}$\cite{BC}. (a) Jets emerging at relativistic speed from the core of active galactic nuclei, from microquasars or gamma-ray bursts. (b) Compact stars and flows around Black Holes. (c) Cosmology. General relativity is necessary only in (b) and (c). In (a) special relativity is sufficient. It is in this context that we introduce a new technique that might render simplification in both mathematical analysis and conceptual understanding. In previous works of the tetrad series we developed a new method to construct new tetrads when second rank antisymmetric fields are present in curved four-dimensional Lorentzian spacetimes. In manuscript \cite{A}$^{,}$\cite{AE} it was the electromagnetic field. In this previous work \cite{A}$^{,}$\cite{AE} we found new tetrads that introduced maximal simplification in the expression of the electromagnetic field, manifestly and covariantly diagonalized at every point the stress-energy tensor for a non-null electromagnetic field and maximally simplified the Einstein-Maxwell equations. In our present case we are dealing with a fluid where the stress-energy tensor can be described by the following equation,

\begin{equation}
T_{\mu\nu}= (\rho + p)\:u_{\mu}\:u_{\nu} + p\:g_{\mu\nu}\ ,\label{SET}
\end{equation}

where $\rho$ is the energy-density of the fluid, $p$ the isotropic pressure and $u^{\mu}$ its four-velocity field, $g_{\mu\nu}$ is the metric tensor. If in addition this fluid has vorticity $\omega_{\mu\nu}$, then we can proceed to build the new tetrads for this particular case following the method developed in \cite{A}$^{,}$\cite{AE}. These new tetrads are going to manifestly and covariantly diagonalize the stress-energy tensor (\ref{SET}) at every spacetime event. We carry out this program in section \ref{fluidtet}. In section \ref{eulervf} using the tetrad found in section \ref{fluidtet}, as an application, we build the Euler observers adapted to the evolution of this kind of spacetimes. Finally, in section \ref{carter} we analyze the Carter-Lichnerowicz equation with our new tetrads as a second application. Specially the barotropic case.

\section{The Fluid Tetrads}
\label{fluidtet}

We introduce the fluid extremal field or the velocity curl extremal field through the local duality transformation given by,

\begin{equation}
\xi_{\mu\nu} = \cos\alpha \:\: u_{[\mu;\nu]} -
\sin\alpha \:\: \ast u_{[\mu;\nu]} ,\label{vef}
\end{equation}

where $\ast u_{[\mu;\nu]} = {1 \over 2}\:\epsilon_{\mu\nu\sigma\tau}\:g^{\sigma\rho}\:g^{\tau\lambda}\:u_{[\rho;\lambda]}$ is the dual tensor of $u_{[\mu;\nu]}$ and the local complexion $\alpha$ is defined through the condition

\begin{equation}
\xi_{\mu\nu}\:\ast \xi^{\mu\nu} = 0 \ .\label{fcond}
\end{equation}

The identity,

\begin{eqnarray}
A_{\mu\alpha}\:B^{\nu\alpha} -
\ast B_{\mu\alpha}\: \ast A^{\nu\alpha} &=& \frac{1}{2}
\: \delta_{\mu}^{\:\:\:\nu}\: A_{\alpha\beta}\:B^{\alpha\beta}  \ .\label{i1}
\end{eqnarray}

which is valid for every pair of antisymmetric tensors in a four-dimensional Lorentzian spacetime \cite{MW}, when applied to the case $A_{\mu\alpha} = \xi_{\mu\alpha}$ and $B^{\nu\alpha} = \ast \xi^{\nu\alpha}$ yields the equivalent condition,

\begin{equation}
\xi_{\mu\rho}\:\ast\xi^{\mu\lambda} = 0 \ .\label{scond}
\end{equation}

The complexion, which is a local scalar, can then be expressed as,

\begin{equation}
\tan(2\alpha) = - \left( u_{[\mu;\nu]}\:g^{\sigma\mu}\:g^{\tau\nu}\:\ast u_{[\sigma;\tau]}\right) \:/\: \left(u_{[\lambda;\rho]}\:g^{\lambda\alpha}\:g^{\rho\beta}\: u_{[\alpha;\beta]}\right) \ .\label{complexion}
\end{equation}

After introducing the new velocity curl extremal field we proceed to write the four orthogonal vectors that are going to become an intermediate step in constructing the tetrad that diagonalizes the stress-energy tensor (\ref{SET}),

\begin{eqnarray}
V_{(1)}^{\alpha} &=& \xi^{\alpha\lambda}\:\xi_{\rho\lambda}\:X^{\rho}
\label{V1}\\
V_{(2)}^{\alpha} &=& \xi^{\alpha\lambda} \: X_{\lambda}
\label{V2}\\
V_{(3)}^{\alpha} &=& \ast \xi^{\alpha\lambda} \: Y_{\lambda}
\label{V3}\\
V_{(4)}^{\alpha} &=& \ast \xi^{\alpha\lambda}\: \ast \xi_{\rho\lambda}
\:Y^{\rho}\ ,\label{V4}
\end{eqnarray}

In order to prove the orthogonality of the tetrad (\ref{V1}-\ref{V4}) it is necessary to use the identity (\ref{i1}) for the case $A_{\mu\alpha} = \xi_{\mu\alpha}$ and $B^{\nu\alpha} = \xi^{\nu\alpha}$, that is,

\begin{eqnarray}
\xi_{\mu\alpha}\:\xi^{\nu\alpha} -
\ast \xi_{\mu\alpha}\: \ast \xi^{\nu\alpha} &=& \frac{1}{2}
\: \delta_{\mu}^{\:\:\:\nu}\:Q\ ,\label{i2}
\end{eqnarray}

where $Q=\xi_{\mu\nu}\:\xi^{\mu\nu}$ is assumed not to be zero. We are free to choose the vector fields $X^{\alpha}$ and $Y^{\alpha}$, as
long as the four vector fields (\ref{V1}-\ref{V4}) are not trivial. It is clear that if our choice for these fields is $X^{\alpha} = Y^{\alpha} = u^{\alpha}$, then the following orthogonality relations will hold,

\begin{eqnarray}
\lefteqn{ g_{\rho\mu}\:u^{\rho}\:V_{(2)}^{\mu} = g_{\rho\mu}\:u^{\rho}\:\xi^{\mu\lambda}\:u_{\lambda} = 0 \label{ortho1} } \\
&&g_{\rho\mu}\:u^{\rho}\:V_{(3)}^{\mu} = g_{\rho\mu}\:u^{\rho}\:\ast\xi^{\mu\lambda}\:u_{\lambda} = 0 \ ,\label{ortho2}
\end{eqnarray}

because of the antisymmetry of the velocity curl extremal field. Then, at the points in spacetime where the set of four vectors (\ref{V1}-\ref{V4}) is not trivial, we can proceed to normalize,

\begin{eqnarray}
\overline{U}^{\alpha} &=& \xi^{\alpha\lambda}\:\xi_{\rho\lambda}\:u^{\rho} \:
/ \: (\: \sqrt{-Q/2} \: \sqrt{u_{\mu} \ \xi^{\mu\sigma} \
\xi_{\nu\sigma} \ u^{\nu}}\:) \label{Uw}\\
\overline{V}^{\alpha} &=& \xi^{\alpha\lambda}\:u_{\lambda} \:
/ \: (\:\sqrt{u_{\mu} \ \xi^{\mu\sigma} \
\xi_{\nu\sigma} \ u^{\nu}}\:) \label{Vw}\\
\overline{Z}^{\alpha} &=& \ast \xi^{\alpha\lambda} \:  u_{\lambda} \:
/ \: (\:\sqrt{u_{\mu}  \ast \xi^{\mu\sigma}
\ast \xi_{\nu\sigma}   u^{\nu}}\:)
\label{Zw}\\
\overline{W}^{\alpha} &=& \ast \xi^{\alpha\lambda}\: \ast \xi_{\rho\lambda}
\: u^{\rho} \: / \: (\:\sqrt{-Q/2} \: \sqrt{ u_{\mu}
\ast \xi^{\mu\sigma} \ast \xi_{\nu\sigma}  u^{\nu}}\:) \ .
\label{Ww}
\end{eqnarray}

In terms of these tetrad vectors (\ref{Uw}-\ref{Ww}) and applying the method developed in manuscript \cite{A}$^{,}$\cite{AE} we can express the velocity curl in its maximal simple form,

\begin{equation}
u_{[\mu;\nu]} = -2\:\sqrt{-Q/2}\:\:\cos\alpha\:\:\overline{U}_{[\alpha}\:\overline{V}_{\beta]} +
2\:\sqrt{-Q/2}\:\:\sin\alpha\:\:\overline{Z}_{[\alpha}\:\overline{W}_{\beta]}\ .\label{VC}
\end{equation}

But these intermediate tetrad (\ref{Uw}-\ref{Ww}) is not the one that diagonalizes the stress-energy tensor. To this end, a new vector field can be defined through the expression,

\begin{eqnarray}
V_{(5)}^{\alpha} = V_{(4)}^{\alpha}\:(V_{(1)}^{\rho}\:u_{\rho}) -  V_{(1)}^{\alpha}\:(V_{(4)}^{\rho}\:u_{\rho}) \ . \label{v5}
\end{eqnarray}

Through the use of the antisymmetry of $\xi_{\mu\nu}$, the condition (\ref{scond}), the identity (\ref{i2}) and the definition of the vectors (\ref{V1}-\ref{V4}), it is simple to prove the following orthogonalities,

\begin{eqnarray}
u_{\mu}\:V_{(5)}^{\mu} = V_{(2)}^{\mu}\:g_{\mu\nu}\:V_{(5)}^{\nu} = V_{(3)}^{\mu}\:g_{\mu\nu}\:V_{(5)}^{\nu} = 0 \ . \label{ortho3}
\end{eqnarray}

Given that $u^{\mu}$, $V_{(2)}^{\mu}$, $V_{(3)}^{\mu}$ and $V_{(5)}^{\mu}$ are orthogonal, we can now proceed to see that these tetrad vectors covariantly and manifestly diagonalize the stress-energy tensor (\ref{SET}) at every spacetime point,

\begin{eqnarray}
u^{\alpha}\:T_{\alpha}^{\:\:\:\beta} &=& -\rho\:u^{\beta}
\label{EV1}\\
V_{(2)}^{\alpha}\:T_{\alpha}^{\:\:\:\beta} &=& p\:V_{(2)}^{\beta}
\label{EV2}\\
V_{(3)}^{\alpha}\:T_{\alpha}^{\:\:\:\beta} &=& p\:V_{(3)}^{\beta}
\label{EV3}\\
V_{(5)}^{\alpha}\:T_{\alpha}^{\:\:\:\beta} &=& p\:V_{(5)}^{\beta}\ .
\label{EV4}
\end{eqnarray}

Finally, we normalize this local tetrad,

\begin{eqnarray}
U^{\alpha} &=& u^{\alpha} \label{U}\\
V^{\alpha} &=& \xi^{\alpha\lambda}\:u_{\lambda} \:
/ \: (\:\sqrt{u_{\mu} \ \xi^{\mu\sigma} \
\xi_{\nu\sigma} \ u^{\nu}}\:) \label{V}\\
Z^{\alpha} &=& \ast \xi^{\alpha\lambda} \: u_{\lambda} \:
/ \: (\:\sqrt{u_{\mu}  \ast \xi^{\mu\sigma}
\ast \xi_{\nu\sigma}   u^{\nu}}\:)
\label{Z}\\
W^{\alpha} &=& \left( V_{(4)}^{\alpha}\:(V_{(1)}^{\rho}\:u_{\rho}) -  V_{(1)}^{\alpha}\:(V_{(4)}^{\rho}\:u_{\rho}) \right) / \: \sqrt{V_{(5)}^{\beta}\:V_{(5)_{\beta}} } \ ,
\label{W}
\end{eqnarray}

where, $V_{(5)}^{\beta}\:V_{(5)_{\beta}} = (V_{(4)}^{\beta}\:V_{(4)_{\beta}})\:(V_{(1)}^{\rho}\:u_{\rho})^{2} + (V_{(1)}^{\beta}\:V_{(1)_{\beta}})\:(V_{(4)}^{\rho}\:u_{\rho})^{2}$.

\section{Application: Euler vector fields}
\label{eulervf}

We are going to proceed in this section in a very similar way to the analogous section in paper \cite{AEO}. We can build tetrads in a similar way to section \ref{fluidtet} but now with no vorticity. We will briefly review this situation and refer the reader of this section to manuscript \cite{ENV} for the detailed presentation of this case. The tetrads found in the case without vorticity are completely analogous to the ones found in the previous section \ref{fluidtet} and that is why we study this case too. We will assume that there is a Killing vector field that will be the basis of an analogous procedure to find a tetrad analogous to the one found in section \ref{fluidtet}, see reference \cite{ENV} for the details. We introduce the equations satisfied by the hypersurface orthogonal \cite{AEO}$^{,}$\cite{ENV}$^{,}$\cite{MC}$^{,}$\cite{RW}$^{,}$\cite{SYO} unit vector fields $n_{\mu}\:n^{\mu} = -1$,

\begin{eqnarray}
n_{\alpha}\:n_{\beta;\gamma} + n_{\beta}\:n_{\gamma;\alpha} +  n_{\gamma}\:n_{\alpha;\beta}
- n_{\alpha}\:n_{\gamma;\beta} - n_{\gamma}\:n_{\beta;\alpha} -  n_{\beta}\:n_{\alpha;\gamma} = 0 \ .\label{hyper}
\end{eqnarray}

We are going to name $\hat{U}^{\mu}$ the Euler unit timelike vector field that satisfies equation (\ref{hyper}). We are going to name the other three  vectors in the new orthonormal tetrad as $\hat{V}^{\mu}$, $\hat{Z}^{\mu}$ and $\hat{W}^{\mu}$. Then, the hypersurface orthogonal vector $\hat{U}^{\mu}$ must satisfy the equation,

\begin{eqnarray}
\hat{U}_{\alpha}\:\hat{U}_{\beta;\gamma} + \hat{U}_{\beta}\:\hat{U}_{\gamma;\alpha} +  \hat{U}_{\gamma}\:\hat{U}_{\alpha;\beta}
- \hat{U}_{\alpha}\:\hat{U}_{\gamma;\beta} - \hat{U}_{\gamma}\:\hat{U}_{\beta;\alpha} -  \hat{U}_{\beta}\:\hat{U}_{\alpha;\gamma} = 0 \ .\label{hyperhat}
\end{eqnarray}

Next, when we project equation (\ref{hyperhat}) using the four tetrad vectors  ($\hat{U}^{\alpha}, \hat{V}^{\alpha}, \hat{Z}^{\alpha}, \hat{W}^{\alpha}$) we get only three meaningful equations,

\begin{eqnarray}
\hat{U}_{[\alpha ; \beta]} \: \hat{V}^{\alpha}\:\hat{Z}^{\beta} &=& 0 \label{VZ}\\
\hat{U}_{[\alpha ; \beta]} \: \hat{V}^{\alpha}\:\hat{W}^{\beta} &=& 0  \label{VW}\\
\hat{U}_{[\alpha ; \beta]} \: \hat{Z}^{\alpha}\:\hat{W}^{\beta} &=& 0   \ . \label{ZW}
\end{eqnarray}

Equations (\ref{VZ}-\ref{ZW}) are three conditions on the vector field $\hat{U}^{\alpha}$. Our intention is to use the tetrad (\ref{U}-\ref{W}) that locally and covariantly diagonalizes the perfect fluid stress-energy tensor, and introduce three local scalars that are going to solve the three equations (\ref{VZ}-\ref{ZW}). To this end, first we perform a rotation on the local plane determined by ($V^{\alpha}, W^{\alpha}$) using the local scalar $\phi$,

\begin{eqnarray}
V^{\alpha}_{(\phi)}  &=& \cos(\phi)\: V^{\alpha} -  \sin(\phi)\: W^{\alpha} \label{VT1} \\
W^{\alpha}_{(\phi)}  &=& \sin(\phi)\: V^{\alpha} +  \cos(\phi)\: W^{\alpha} \label{WT1} \ .
\end{eqnarray}

Second, we perform another local rotation in the plane ($Z^{\alpha}, W^{\alpha}_{(\phi)}$) by the local angle $\varphi$,

\begin{eqnarray}
Z^{\alpha}_{(\varphi)}  &=& \cos(\varphi)\: Z^{\alpha} -  \sin(\varphi)\: W^{\alpha}_{(\phi)} \label{ZT1} \\
W^{\alpha}_{(\varphi)}  &=& \sin(\varphi)\: Z^{\alpha} +  \cos(\varphi)\: W^{\alpha}_{(\phi)} \label{WT2} \ .
\end{eqnarray}

Finally a boost by the local angle $\psi$ in the plane ($U^{\alpha}, W^{\alpha}_{(\varphi)}$),

\begin{eqnarray}
\hat{U}^{\alpha}  &=& \cosh(\psi)\:U^{\alpha}  +  \sinh(\psi)\:W^{\alpha}_{(\varphi)}  \label{UWf} \\
\hat{W}^{\alpha}  &=& \sinh(\psi)\:U^{\alpha}  +  \cosh(\psi)\:W^{\alpha}_{(\varphi)}  \label{UWs} \ .
\end{eqnarray}

Three local scalars ($\phi$, $\varphi$, $\psi$) become through these succession of local Lorentz transformations in three local variables that are going to be the solution to the system (\ref{VZ}-\ref{ZW}). The final orthonormal tetrad that has as a timelike vector field $\hat{U}^{\alpha}$ the hypersurface orthogonal vector field that will function as an input for our evolution algorithms is given by,

\begin{eqnarray}
\hat{U}^{\alpha}  &=& \cosh(\psi)\:U^{\alpha}  +  \sinh(\psi)\:W^{\alpha}_{(\varphi)}  \label{SFU} \\
\hat{V}^{\alpha} &=& V^{\alpha}_{(\phi)} \label{SFV} \\
\hat{Z}^{\alpha}  &=& Z^{\alpha}_{(\varphi)}\label{SFZ} \\
\hat{W}^{\alpha}  &=& \sinh(\psi)\:U^{\alpha}  +  \cosh(\psi)\:W^{\alpha}_{(\varphi)}  \label{SFW} \ .
\end{eqnarray}

The algorithm would not work if the vector that involves the three local Lorentz transformations and therefore the three local scalars $(\phi, \varphi, \psi)$, were not $\hat{U}^{\alpha}$. If we would have considered Lorentz transformations only involving the original vectors $(V^{\alpha}, Z^{\alpha}, W^{\alpha})$ then we would only have produced combinations of the original equations (\ref{VZ}-\ref{ZW}) and since these can be algebraically decoupled, we would not have introduced any new information. It is through the inclusion of the three local scalars $(\phi, \varphi, \psi)$ inside the derivatives of the vector $\hat{U}^{\alpha}$ that we get equations (\ref{VZ}-\ref{ZW}) to be meaningful. Next, we contract the tetrad vectors $(\hat{U}^{\alpha}, \hat{V}^{\alpha}, \hat{Z}^{\alpha}, \hat{W}^{\alpha})$ with the stress-energy tensor (\ref{SET}),

\begin{eqnarray}
\hat{U}^{\alpha}\:T_{\alpha}^{\:\:\:\beta} &=& -\rho\:\cosh(\psi)\:U^{\beta} + p\:\sinh(\psi)\:W^{\beta}_{(\varphi)} \label{EV1}\\
\hat{V}^{\alpha}\:T_{\alpha}^{\:\:\:\beta} &=& p\:\hat{V}^{\beta} \label{EV2}\\
\hat{Z}^{\alpha}\:T_{\alpha}^{\:\:\:\beta} &=& p\:\hat{Z}^{\beta} \label{EV3}\\
\hat{W}^{\alpha}\:T_{\alpha}^{\:\:\:\beta} &=& -\rho\:\sinh(\psi)\:U^{\beta} + p\:\cosh(\psi)\:W^{\beta}_{(\varphi)}\ .\label{EV4}
\end{eqnarray}

Therefore, the only non-zero components of the stress-energy tensor in terms of the new tetrad are,

\begin{eqnarray}
\hat{U}^{\alpha}\:T_{\alpha}^{\:\:\:\beta}\:\hat{U}_{\beta} &=& \rho\:\cosh^{2}(\psi) + p\:\sinh^{2}(\psi) \label{SE00F}\\
\hat{V}^{\alpha}\:T_{\alpha}^{\:\:\:\beta}\:\hat{V}_{\beta} &=& p \label{SE11F}\\
\hat{Z}^{\alpha}\:T_{\alpha}^{\:\:\:\beta}\:\hat{Z}_{\beta} &=& p \label{SE22F}\\
\hat{W}^{\alpha}\:T_{\alpha}^{\:\:\:\beta}\:\hat{W}_{\beta} &=& \rho\:\sinh^{2}(\psi) + p\:\cosh^{2}(\psi) \label{SE33F}\\
\hat{U}^{\alpha}\:T_{\alpha}^{\:\:\:\beta}\:\hat{W}_{\beta} &=& \frac{(\rho + p)}{2}\:\sinh(2\psi)\ .\label{SE03F}
\end{eqnarray}

By performing the three local Lorentz transformations in our algorithm we have the following result. First, we ended up with a new local tetrad that adds only one off-diagonal component to the stress-energy tensor, the minimum possible. Second, we found the Euler hypersurface orthogonal congruence. We have found an algorithm that provides both a hypersurface orthogonal congruence and a maximum simplification of the stress-energy tensor given that the tetrad that diagonalized the tensor underwent three Lorentz transformations. When we take the limit $\psi \rightarrow 0$ it can be readily seen from expressions (\ref{SE00F}-\ref{SE03F}) that we recover the results for the old tetrad that diagonalizes the stress-energy tensor.

\section{Application: Carter-Lichnerowicz equation}
\label{carter}

We start by introducing the Carter-Lichnerowicz equation. For instance,

\begin{eqnarray}
u^{\mu}\:[\frac{\partial (h\:u_{\alpha})}{\partial x^{\mu}} - \frac{\partial (h\:u_{\mu})}{\partial x^{\alpha}}]  = T\: \frac{\partial \overline{s}}{\partial x^{\alpha}} \label{carterlicheq} \ ,
\end{eqnarray}

where $h$ is the enthalpy per baryon, and $\overline{s}$ is the entropy per baryon, following the notation in \cite{EG}. It is evident through all our previous work in references \cite{A}$^{,}$\cite{AE}$^{,}$\cite{AEO} that we can apply our tetrad construction system to the object $\omega_{\mu\alpha} = \frac{\partial (h\:u_{\alpha})}{\partial x^{\mu}} - \frac{\partial (h\:u_{\mu})}{\partial x^{\alpha}}$. Let us call the tetrads obtained through this process,

\begin{eqnarray}
\tilde{U}^{\alpha} &=& \xi^{\alpha\lambda}\:\xi_{\rho\lambda}\:u^{\rho} \:
/ \: (\: \sqrt{-Q/2} \: \sqrt{u_{\mu} \ \xi^{\mu\sigma} \
\xi_{\nu\sigma} \ u^{\nu}}\:) \label{UCL}\\
\tilde{V}^{\alpha} &=& \xi^{\alpha\lambda}\:u_{\lambda} \:
/ \: (\:\sqrt{u_{\mu} \ \xi^{\mu\sigma} \
\xi_{\nu\sigma} \ u^{\nu}}\:) \label{VCL}\\
\tilde{Z}^{\alpha} &=& \ast \xi^{\alpha\lambda} \:  u_{\lambda} \:
/ \: (\:\sqrt{u_{\mu}  \ast \xi^{\mu\sigma}
\ast \xi_{\nu\sigma}   u^{\nu}}\:)
\label{ZCL}\\
\tilde{W}^{\alpha} &=& \ast \xi^{\alpha\lambda}\: \ast \xi_{\rho\lambda}
\: u^{\rho} \: / \: (\:\sqrt{-Q/2} \: \sqrt{ u_{\mu}
\ast \xi^{\mu\sigma} \ast \xi_{\nu\sigma}  u^{\nu}}\:) \ .
\label{WCL}
\end{eqnarray}

Similarly to previous applications we have $\xi_{\mu\nu} = \cos\alpha \:\: \omega_{\mu\nu} - \sin\alpha \:\: \ast \omega_{\mu\nu}$, $Q = \xi_{\mu\nu}\:\xi^{\mu\nu}$ and $\omega_{\mu\nu} = -2\:\sqrt{-Q/2}\:\:\cos\alpha\:\:\tilde{U}_{[\mu}\:\tilde{V}_{\nu]} +
2\:\sqrt{-Q/2}\:\:\sin\alpha\:\:\tilde{Z}_{[\mu}\:\tilde{W}_{\nu]}$. We also have $\tan(2\alpha) = - \omega_{\mu\nu}\:\ast \omega^{\mu\nu} / \omega_{\lambda\rho}\:\omega^{\lambda\rho}$. It is clear to see that $u^{\mu}\:\tilde{V}_{\mu} = u^{\mu}\:\tilde{Z}_{\mu} = 0$ by construction. Then, $u^{\mu}\:\omega_{\mu\nu} = -\:\sqrt{-Q/2}\:\:\cos\alpha\:(\:u^{\mu}\:\tilde{U}_{\mu})\:\tilde{V}_{\nu} -
\:\sqrt{-Q/2}\:\:\sin\alpha\:\:(u^{\mu}\:\tilde{W}_{\mu})\:\tilde{Z}_{\nu}$. There we have first, the most simplified expression of the tensor $\omega_{\mu\nu}$ in terms of the new tetrad $(\tilde{U}^{\mu}, \tilde{V}^{\mu}, \tilde{Z}^{\mu}, \tilde{W}^{\mu})$, and a second simplification since only two terms survive the contraction $u^{\mu}\:\omega_{\mu\nu}$. The new expression for the Carter-Lichnerowicz equation becomes,

\begin{eqnarray}
-\:\sqrt{-Q/2}\:\:\cos\alpha\:(\:u^{\mu}\:\tilde{U}_{\mu})\:\tilde{V}_{\nu} -
\:\sqrt{-Q/2}\:\:\sin\alpha\:\:(u^{\mu}\:\tilde{W}_{\mu})\:\tilde{Z}_{\nu} = T\: \frac{\partial \overline{s}}{\partial x^{\alpha}} \label{newcarterlicheq} \ .
\end{eqnarray}

A particular case of relevance would be the isentropic case or barotropic fluid. It is specially important for cold dense matter in white dwarfs and neutron stars. The equation would be in this case,

\begin{eqnarray}
-\:\sqrt{-Q/2}\:\:\cos\alpha\:(\:u^{\mu}\:\tilde{U}_{\mu})\:\tilde{V}_{\nu} -
\:\sqrt{-Q/2}\:\:\sin\alpha\:\:(u^{\mu}\:\tilde{W}_{\mu})\:\tilde{Z}_{\nu} = 0 \label{barotropic} \ .
\end{eqnarray}

In terms of our new tetrads we can deduce the following. We have two possibilities. First since the vectors $(\tilde{V}^{\mu}, \tilde{Z}^{\mu})$ are orthogonal since the tetrad (\ref{UCL}-\ref{WCL}) is orthonormal, we must have $\:u^{\mu}\:\tilde{U}_{\mu} = \:u^{\mu}\:\tilde{W}_{\mu} = 0$. But then, it is not possible to normalize the vectors (\ref{UCL}-\ref{WCL}). It must be that the two conditions $u_{\mu} \: \xi^{\mu\sigma} \: \xi_{\nu\sigma} \: u^{\nu} = 0$ and $u_{\mu}
\ast \xi^{\mu\sigma} \ast \xi_{\nu\sigma} \: u^{\nu} = 0$ hold at every point from the outset, before we start our tetrad construction process. The second possibility is that $Q = \xi_{\mu\nu}\:\xi^{\mu\nu} = 0$ at every point. Therefore, reducing the barotropic case to a pure simplified geometric and algebraic problem in either case.

\section{Conclusions}

We believe that this new tetrad can be used and bring about simplification in the analysis of astrophysical relativistic problems where vorticity is present, for instance through the Carter-Lichnerowicz equation, see section \ref{carter} and reference \cite{EG}. To this end we have to observe on one hand that equation (\ref{VC}) expresses the velocity curl in its maximal simple form, and on the other hand that if we compare the tetrad set (\ref{Uw}-\ref{Ww}) with the tetrad set (\ref{U}-\ref{W}) we notice that $\overline{V}^{\mu} = V^{\mu}$ and $\overline{Z}^{\mu} = Z^{\mu}$. In section \ref{carter} we studied the application of these new tetrads to the Carter-Lichnerowicz equation for several situations like the barotropic fluid, for instance. This case is relevant because of cold dense matter in white dwarfs and neutron stars. This general algorithm presented in this paper, provided that solutions exist, reduces a local diagonalization process into a covariant local algebraic process. By means of tetrad vectors that make sense from the point of view of the available geometric structures present in these perfect fluid environments. Otherwise, the stress-energy tensor would have to be diagonalized at every point, blindly, without any tetrad vectors, thus preventing a clear visualization of the geometrical underlying structures present in this problem. Relativistic fluid dynamics addresses astrophysical phenomena directly related to sources of gravitational waves \cite{AC}$^{,}$\cite{F} where this new tetrad finds applications. The fundamental question would be if any advantage is gained through the use of tetrad (\ref{U}-\ref{W}) with respect to standard local diagonalization of the stress-energy tensor. We can argue from the outset that we would be using available background meaningful structures in order to simplify the geometrical visualization and mathematical analysis as can be noticed through equation (\ref{VC}). Second, as another application, these new tetrads would become relevant when evolving spacetimes through the use of Cauchy surfaces and specially in order to build Euler observers as can be readily seen from the work developed in manuscript \cite{AEO} and section \ref{eulervf}. As the analysis is analogous for the case without vorticity and the presence of a Killing vector field we briefly reviewed the construction of Euler observers studied in reference \cite{ENV}. The tetrads found in the case with no vorticity and a Killing vector field are completely analogous to the tetrads with vorticity and that is why we pay attention to this case, see reference \cite{ENV} for the details. Three local Lorentz transformations of the tetrad (\ref{U}-\ref{W}) allow for the construction of Euler observers that in turn simplify the Cauchy evolution problem \cite{JWY}$^{-}$\cite{GC}. We quote from \cite{F} ``The numerical investigation of many interesting astrophysical processes involving neutron stars, such as the rotational evolution of proto-neutron stars (which can be affected by a dynamical bar mode instability and by the Chandrasekhar-Friedman-Schutz instability) or the gravitational radiation from unstable pulsation modes or more importantly, from the catastrophic coalescence and merger of neutron star compact binaries, requires the ability of accurate, long-term hydrodynamical evolutions employing relativistic gravity. These scenarios are receiving increasing attention in recent years''. On the other hand, following the ideas in \cite{CW}$^{,}$\cite{B}$^{,}$\cite{BB} we can readily see that ``inertia here'' is produced by the energy-density, pressure, vorticity and gravity itself, ``there''. We can visualize all this through the fluid differential equations in a curved spacetime, and the tetrads (\ref{U}-\ref{W}) themselves. Matter contributes through the energy-density, pressure and vorticity to define the local tetrads ``here''. The velocity curl is explicitly involved in the tetrad vectors construction through the velocity curl extremal field that we also called fluid extremal field. Energy-density, pressure and vorticity define the gravitational field through the solutions to the differential equations. Gravity also produces ``inertia here'' through the non-linearities of the differential equations where gravity is a source to gravity itself. The metric tensor is also directly involved in the construction of the tetrad vectors. We quote from \cite{B} ``Historically, dynamics was bedevilled from its beginning by the invisibility of space and time. Newton (1686) championed the view that space and time, although invisible, do exist and provide the arena within which motion occurs. Leibnitz (1716) argued that there is no such thing as absolute space but only the relative configurations of simultaneously existing bodies and that time is merely the succession of such instantaneous configurations and not something that flows quite independently of the bodies in the universe and their motion. What Leibnitz was advocating was that dynamics should be based exclusively on observable elements; it should not contain elements that are not in principle observable. This, of course, was Mach's standpoint too (Mach 1872), which Einstein (1916) adopted wholeheartedly when developing General Relativity''.

%The presence of the four-velocity field in the tetrad construction is of physical relevance as far as it is necessary to define the tetrad that diagonalizes the stress-energy tensor. However, it is a gauge choice. A transformation like $u^{\alpha} \rightarrow u^{\alpha}+\partial^{\alpha}\Lambda$ where $\Lambda$ is a scalar field, would generate a tetrad rotation on the planes generated by $(U^{\alpha}, V^{\alpha})$ on one hand, and $(Z^{\alpha}, W^{\alpha})$ on the other hand. The new vectors would be a tetrad but not the one that diagonalizes the stress-energy tensor. We refer the reader to the theorems proved in a previous paper of the tetrad series, in the section group isomorphisms \cite{A}.

%\bibliography{your-bib-file} % place the references here.

\end{document}